# On the room-temperature aging effects in $YBa_2Cu_3O_{6+\delta}$

**A. V. Fetisov, G. A. Kozhina, S. Kh. Estemirova, V. Ya. Mitrofanov**

*Institute of Metallurgy, Ural Branch of the Russian Academy of Sciences,101 Amundsen str., 620016 Ekaterinburg, Russia*

Correspondence should be addressed to Andrew Fetisov, fetisovav@mail.ru

**Abstract**

The aging effects in $YBa_2Cu_3O_{6+\delta}$ have been investigated on the low-speed quenched samples in contrast to other similar studies where the high-speed quenched samples were examined. In the framework of the investigation, XPS analysis has been conducted for the samples stored after quenching for the specified times $\tau$ and a detailed experimental dependence of the $YBa_2Cu_3O_{6+\delta}$ lattice parameters *a*, *b*, and *c* on both $\delta$ and $\tau$ has been obtained. The regime of low-speed quenching results in a threefold increase in the aging effect magnitude. Particular cases of the $c(\delta; \tau)$ dependence are well described by a second-order polynomial with applying different coefficients for different $\tau$. The behavior *c* has been explained on the basis of a simple model of the Coulomb interaction between the $CuO_2$ and $CuO_\delta$ structural planes, in which the hole charge transfer from $CuO_\delta$ to $CuO_2$ takes place providing the standard level of $T_c$ in $YBa_2Cu_3O_{6+\delta}$ (this level is characterized by a maximum $T_c$ of 93 K at $\delta = 0.9$). It has been demonstrated that the $c(\delta; \tau)$ dependence is a good alternative to assessing the hole concentration in $YBa_2Cu_3O_{6+\delta}$. By analyzing the $c(\delta; \tau)$ experimental dependence as well as the data on the hole localization in the pairs Cu–O obtained by XPS, the authors have made a conclusion about the nature of observed aging effects. The latter are likely to be related to the transition of the Cu $3d^9L^{-1}$ electron configuration to Cu $3d^8$ for copper in the basal plane of $YBa_2Cu_3O_{6+\delta}$ at RT.

## 1. Introduction

It is well known that the superconducting transition temperature $T_c$ of $YBa_2Cu_3O_{6+\delta}$ depends significantly on the oxygen parameter $\delta$, varying from 0 K (at $\delta \approx 0$) to 93 K (at $\delta \approx 0.9$). At a constant $\delta$ the temperature $T_c$ is believed to be quite stable and its small changes occurring over time are associated with the processes of chemical degradation [1, 2].

It has been established, however, that the $YBa_2Cu_3O_{6+\delta}$ samples ($\delta < 0.7$) just quenched after high-temperature oxidation to room temperature (RT) have the lattice parameters and $T_c$ considerably different from the values found for samples stored for a long time $\tau$ (hereafter post-quench time or $\tau$) [3–9]. Eventually, these values reach the level specific for given $\delta$. To illustrate that, let us consider results presented by some groups of investigators. So, in the study of Jorgensen *et al.* [6], the continuous growth of $T_c$ from 0 to 20 K and the fall of the parameter $c$ were detected for a few days after the $YBa_2Cu_3O_{6.41}$ had been quenched. The authors found these changes to occur by rapidly damped law. The time constant was found to be 386 minutes and the total reduction of the parameter $c$ was 0.04%. In the study of Kircher *et al.* [8], the 15 hour-growth of $T_c$ from 16 K (at $\tau = 1$ h) to 35 K for the $YBa_2Cu_3O_{6.35}$ sample and from 47 to 55 K for the $YBa_2Cu_3O_{6.45}$ was observed. The elevation of $T_c$ was accompanied by increasing the $Cu^+$ concentration. The relaxation of the lattice parameter $c$ occurring in the sample $YBa_2Cu_3O_{6.25}$ at RT was studied by Shaked *et al.* [5]. The resultant decrease in this parameter of ~ 0.0015 Å, reaching logarithmically with the relaxation time of ~ 760 min, was recorded. All these effects are often called *room-temperature aging effects*. The kinetics of the aging effect in $YBa_2Cu_3O_{6+\delta}$ with small deviations in the storage temperature from RT was studied by Veal *et al.* [7]. The strong dependence of the relaxation rate on temperature observed in this investigation indicated that processes underlying the aging effect are of thermally activated nature.

Typically, the changes in $T_c$ and $c$ observed after quenching $YBa_2Cu_3O_{6+\delta}$ are explained by the ordering processes proceeding in the incomplete oxygen sublattice of the $CuO_\delta$ plane at RT, Fig. 1(a). Herewith, it is generally accepted that $T_c$ is greater, when the degree of oxygen ordering is higher [4–15]. In the general case, the oxygen ordering degree can be determined as the part of oxygen in the $CuO_\delta$ plane belonging to the chain fragments of type I, see Fig. 1(b), in which the copper ions are square-planar coordinated by oxygen and their formal valence state is Cu(III). Such fragments are considered [5, 14–18] as a source of the hole formation unlike other types of fragments: II and III, see. Fig. 1(b). Meanwhile, the hole concentration in $CuO_2$ planes [Fig. 1(a)], where superconductivity is realized, depends directly on the hole concentration in $CuO_\delta$ through the mechanism of charge transfer [14, 17]. Using this model it is suggested that aging effects are controlled by the kinetics of an ortho-I $\rightarrow$ ortho-II transition occurring in $YBa_2Cu_3O_{6+\delta}$ at RT, where the ortho-I phase (see Fig. 1) likely to contain few copper ions in the chain fragments of type I is formed during high-temperature oxidation and the ortho-II phase (with alternating Cu-O-Cu and Cu-$V_o$-Cu chains along the *a*-direction of $CuO_\delta$ planes) arises at RT as thermodynamically stable superstructure and has a lot more Cu(III) ions. In turn, the mechanism of the time-dependent decrease in the parameter $c$ is still under question. This problem was touched in [5] where the authors supposed the lattice parameters to be related to the charge transfer occurring between the $CuO_\delta$ and $CuO_2$ planes (as a consequence of oxygen ordering), like in the case of $T_c$.

To our knowledge, however, there are no direct evidences linking the aging effects with the realization of low-temperature superstructure orderings. Furthermore, more recent investigations provide convincing examples which are in poor agreement with this correlation. So, the kinetics of the ortho-I → ortho-II transition studied by X-ray diffraction [19, 20] turned out to be characterized by the relaxation time (about 100 hours at 70–75°C) which is much longer than that typical for the kinetics of aging effects. Besides, in studies where comparative analysis of $T_c$ was carried out on the ortho-II and ortho-I based samples with the same $\delta$ [13, 21], a difference in $T_c$ was found to be not more than 5 K for the entire range of compositions examined. By annealing samples in the range 25–120 °C (in order to vary their ortho-II ordering degree), a maximum difference in $T_c$ amounted to only 2 K ($51 \leq T_c \leq 53$ K) [22]. Finally, the enthalpy of the disproportionation reaction

$$Cu^{2+} \rightarrow Cu^{+} + Cu^{3+}, \qquad (1)$$

which controls the aging effects [according to Knizhnik *et al.* [3], this enthalpy is about 10 kJ/(mol $YBa_2Cu_3O_{6+\delta}$)] is too large to pertain to the "driving force" of the ortho-I → ortho-II transition (interactions playing a leading role in this transition are of order 1 kJ/(mol $YBa_2Cu_3O_{6+\delta}$) [23]). This is the thermodynamic support for the findings [13, 21] that proper values of $T_c$ can be achieved without necessity for the ortho-II structure to be formed.

(a)              (b)

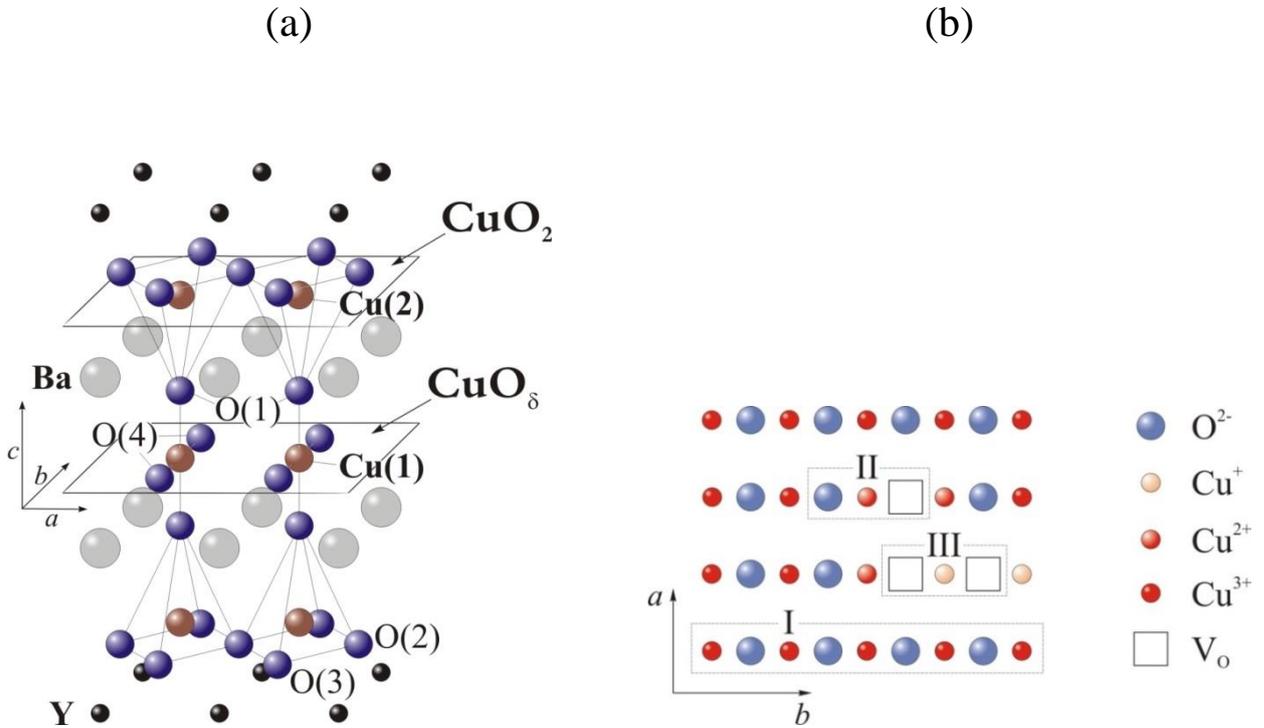

Fig. 1. (a) The crystal lattice of $YBa_2Cu_3O_{6+\delta}$; (b) Schematic illustration of the $CuO_\delta$ plane in $YBa_2Cu_3O_{6+\delta}$ (*ab0*), within which ordering of oxygen ions occurs (Cu-O-Cu chains extended along the *b*-axis). Shown structure is named ortho-I phase. Highlighted fragments I, II and III correspond to different types of copper ion coordination. $V_o$ denotes the oxygen vacancy.

Thus, the nature of the aging effects cannot be considered as fully clarified. The problem is complicated by deficiency of the information on changing concentration of holes with time at RT. Currently, available estimates are not numerous and do not agree well with each other.

In order to suggest a new explanation for the room-temperature aging effects, XPS data about the chemical state of oxygen in $YBa_2Cu_3O_{6+\delta}$ and the lattice parameters vs $\delta$ detailed experimental dependences have been obtained for different values of post-quench time $\tau$. We based on a reasonable assumption that the change of the lattice parameter $c$ occurring over time must directly reflect the process of the charge transfer between the planes $CuO_\delta$ and $CuO_2$. This assumption allows analyzing the $\delta$- and $\tau$-dependent hole concentration in the $CuO_2$ plane and juxtaposing it with the chemical state of oxygen in $YBa_2Cu_3O_{6+\delta}$ showing clearly the presence of holes on a part of oxygen at $\tau \to 0$ and their disappearance later on. As a result, we have suggested a new mechanism for changing the hole concentration in the $CuO_2$ plane independent from oxygen ordering in $YBa_2Cu_3O_{6+\delta}$.

## 2. Experiment details

$YBa_2Cu_3O_{6+\delta}$ was synthesized from the oxides $Y_2O_3$, CuO and barium carbonate $BaCO_3$, which were mechanically mixed and annealed at 950°C for 100 hours. As a result, single phase product of the tetragonal structural modification was formed. To obtain samples with different oxygen content, portions of the synthesized material (by ~ 5–6 g) were oxidized at different temperatures from the interval 470–950°C in air atmosphere and then quenched. The duration of oxidative annealing ranged from 1.5 to 2.5 h depending on the temperature of the process. The uniform distribution of oxygen throughout the material volume was achieved by repeated oxidation of the samples at the same thermodynamic parameters (t, $p_{O_2}$). However, the heat treatment was conducted now in crucibles partially covered with lids. To avoid additional oxidation of the samples during their cooling, crucibles with samples were fully closed before the quenching procedure. Material for research was taken from the central part of the sintered sample.

The oxygen content was monitored by the change in mass of a small portion of the sample during additional 1 h oxidation at 470°C in air. Using the values of $\Delta m$, the initial composition of a sample was determined by the formula:

$$6 + \delta = 6.9 - \left[ \frac{2.43 \cdot m}{\Delta m} \right], \qquad (2)$$

where 6.9 is the oxygen content in $YBa_2Cu_3O_{6+\delta}$ corresponding to the conditions of its additional oxidative annealing (t = 470°C, $p_{O_2}$ = 21 kPa, $\tau$ = 1 h); $m$ - mass of the sample. There was a good correlation between $\delta$ values calculated by Eq. (2) and conditions of oxidative annealing (at least for samples with orthorhombic structure), the relationship between which was established in the study on thermodynamic equilibrium of $YBa_2Cu_3O_{6+\delta}$ with the gas phase [24].

X-ray diffraction (XRD) was performed on a Shimadzu XRD-7000 diffractometer by scanning in the range $2\theta = 20–60°$. The X-ray photoelectron spectroscopy (XPS) studies were carried out on an electron spectrometer based on a Multiprobe Compact vacuum system and an EA-125 energy analyzer (Omicron, Germany). The energy scale of the spectrometer was calibrated using the Au $4f_{7/2}$, Ag $3d_{5/2}$, and Cu $2p_{3/2}$ lines. The peak positions were corrected, taking into account the "charging" of the samples when exposed to radiation, using the lowest-energy carbon line C 1$s$ (285.0 eV).

The superconducting transition temperature $T_c$ of the samples was determined using the low-temperature magnetic measurements (LTMM) in a vibrating sample magnetometer (VSM) Cryogenic CFS-9T-CVTI, measuring the in-field cooling magnetization at 50 Oe in the temperature range 5–150 K. Cooling the samples to 5 K and subsequent heating to room temperature was performed with a rate of 1°C/min.

Every sample quenched after oxidative annealing was examined by the XPS, XRD, and LTMM methods according to the scheme: (1) XRD survey in 0.30±0.03 h after quenching (hereafter $\tau = 0.3$ h), (2) XPS surveys in 2.1 h (hereafter $\tau = 2$ h), (3) XRD and XPS surveys in 5.1 h (hereafter $\tau = 5$ h), (4) XPS surveys in 24.1 h (hereafter $\tau = 1$ day), and (5) XRD, XPS and LTMM surveys in 15±2 days after quenching (hereafter $\tau = 15$ days). The middle of the survey duration was taken as the moment of survey in the calculation of $\tau$.

## 3. Experimental results

### 3.1. XPS analysis

Our studies were conducted on the $YBa_2Cu_3O_{6+\delta}$ sample with $\delta = 0.65$ and were focused on the oxygen peak O 1$s$. Fig. 2 shows the set of O 1$s$ spectra obtained at specified post-quench time $\tau$. Since the observable spectral changes occurring over time were small, in order to visualize them we plotted difference spectra (Fig. 2 shows an example of such a spectrum): the spectrum corresponding to the beginning of the survey cycle ($\tau = 2$ h) was subtracted from the spectra obtained at the subsequent investigation stages. This was done after normalization of the spectra. The spectral changes occurring over time included a shift of a small component of the O 1$s$ spectrum from the initial position near 530 eV to ~528 eV, which was clearly seen as dips and maxima (having approximately equal areas) on difference spectra corresponding to $\tau = 5$ h ÷ 15 days. These data were used to create a model for analyzing the original experimental data obtained: the initial oxygen spectra were represented by the convolution of three stationary peaks and one moving peak (under the arrow in Fig. 2), the position of which depends on the post-quench time.

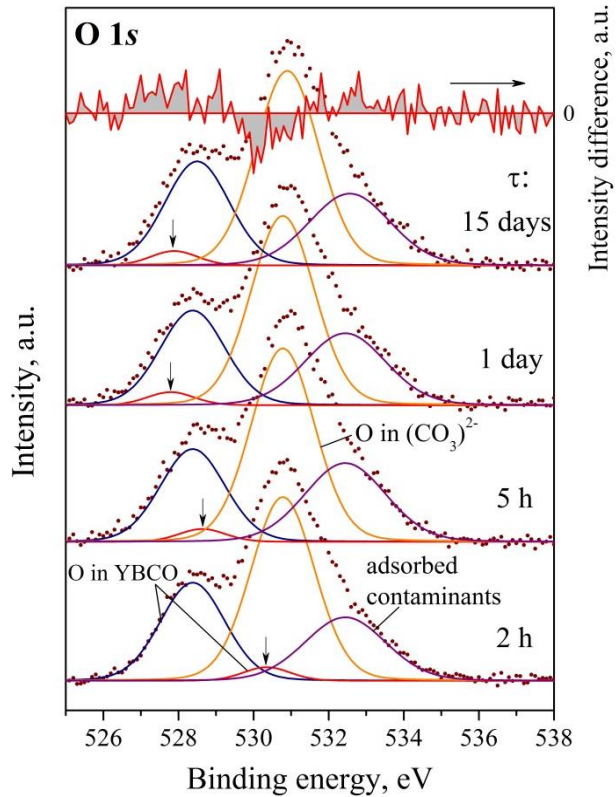

Fig. 2. XPS O 1*s* spectra of $YBa_2Cu_3O_{6.65}$ stored after quenching for the specified times τ. *The right axis* (the top graph): the difference between 15 days- and 2 h- spectra (for details see the text); resulting spectrum is highlighted by the grey shading.

According to [34], the O 1*s* signal inherent to the oxygen ions in $YBa_2Cu_3O_{6+\delta}$ (existing predominantly as $O^{2-}$ species) is recorded at 528.5 eV. Consequently, the stationary lowest-energy peak in the O 1*s* spectra in Fig. 2 is associated with the main phase of the sample. Also, the extreme low-energy position (527.8±0.1 eV) of the moving peak, obviously, corresponds to the main valence state of oxygen in $YBa_2Cu_3O_{6+\delta}$, with excess electron density localized on this anion. But then, the initial position of this peak (530.3 eV), should be related to the oxygen which has a relatively diminished electron density in the valence shell. According to the K. Zigban conception [26], the shift of peak in the O 1*s* spectra (the so-called *chemical shift*) could be interpreted as a manifestation of the electron-charge transfer from ligand to metal. As regards the quantitative indicators, calculations presented by Moulder *et al.* [27] show that such chemical shift as observed in Fig. 2 reflects an increase in the negative charge on oxygen from -0.83 to -1.02 esu and this increase is very large. In turn, the moving peak area, which can be calculated from the dip on difference spectra, is about 1/10 of the main signal for $YBa_2Cu_3O_{6.65}$ (peak at 528.4 eV). This ratio corresponds to the fraction of oxygen in $YBa_2Cu_3O_{6.65}$ which is related to the $CuO_\delta$ plane [$\delta/(6+\delta) = 0.65/6.65 \approx 0.1$]. Hence, the moving peak is more likely to be the signal from this oxygen [oxygen O(4) in Fig. 1(a)].

As for other peaks in the O 1*s* spectra, a peak at 530.9 eV is associated with carbonate group $(CO_3)^{2-}$ and a peak at 532.4 eV – with organic contaminants adsorbed on the surface, which is typical for XPS studies [28].

## 3.2. X-ray analysis

The dependences of the lattice parameters $a$, $b$, and $c$ (the axis $c$ is directed along the normal to the structural planes of $YBa_2Cu_3O_{6+\delta}$) on the oxygen content and the post-quench time $\tau$ are shown in Fig. 3. It is seen that non-monotonic particular dependence $c(\delta)$ obtained for $\tau = 0.3$ h passes into smooth curves later on ($\tau \geq 5$ h). The course of the $a,b(\delta, \tau)$ curves reflects the existence of the structural tetra→ortho transition occurring in $YBa_2Cu_3O_{6+\delta}$ at $\delta \approx 0.32$ [13, 16, 18]. It can be seen that the $\tau$ dependence of $a$ and $b$ is rather weak at $\tau \leq 5$ h, however, at $\tau > 5$ h the $a(\delta)$ curve is noticeably changed that looks like the behavior of the parameter $c$. Since the total duration of the aging processes has never exceeded 15 days (e.g. see Refs. [4–8]), this implies that datasets "15 days" in Fig. 3 may be considered as finally formed dependences (equilibrium at RT).

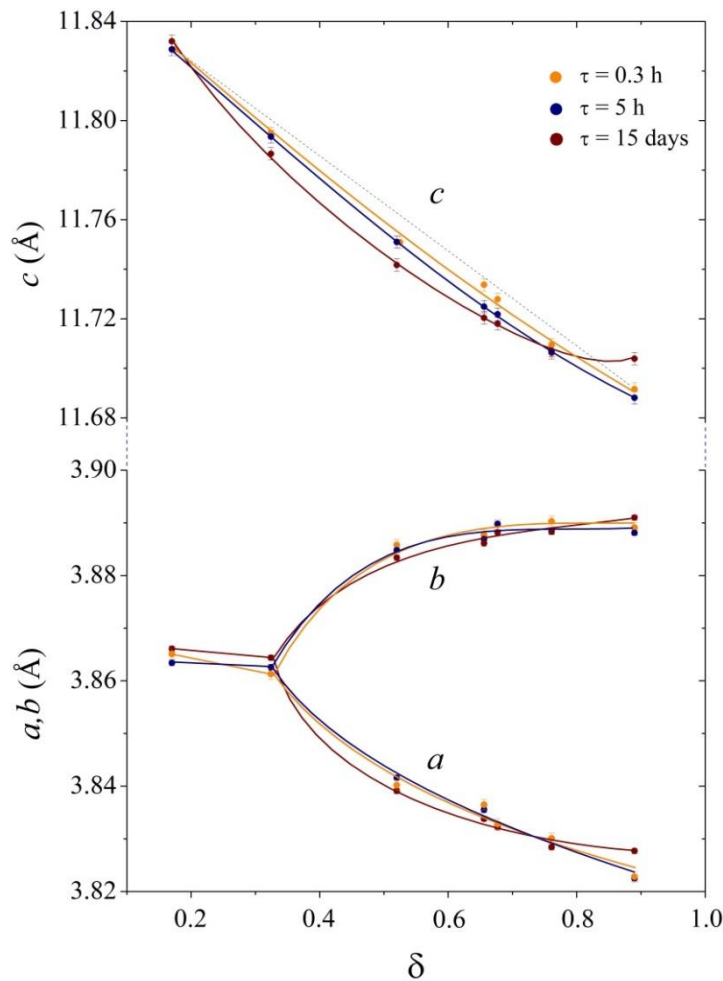

Fig. 3. The experimental dependences of lattice parameters $c$, $a$, and $b$ on excess oxygen content $\delta$ and the post-quench time for $YBa_2Cu_3O_{6+\delta}$. The curves are guides to the eye. The dashed line in the top panel is straight; it is aimed to highlight the curvature of the other curves.

It should be noted that the study of the lattice parameters relaxation on quenched $YBa_2Cu_3O_{6+\delta}$ was also carried out in other works, however, these studies touched on only a few particular cases of δ. Therefore, those data do not provide an overall picture of the lattice parameter changes occurring in the coordinates (δ, τ). For example, the authors of [5, 6] found that the parameter $c$ was only reduced with increasing τ. But, we have revealed the area (δ ≥ 0.76) where this parameter either increases or remains constant (Fig. 4). Considering the dependence $c(δ, τ)$ for wide ranges of δ and τ we can see that storing of $YBa_2Cu_3O_{6+δ}$ at RT results in changing $c$ with definite regularities. So, both the tilt angle with respect to the δ axis and curvature of the $c(δ)$ dependence grow with increasing τ.

Analyzing the magnitude of the aging effect obtained by us, we surprisingly found that it significantly differs from literature data. That is well illustrated in Fig. 4 where the magnitude of changes of the parameter $c$ observed in [5, 6] is compared with that obtained in the present work for the same values of δ. One can see that in our case the aging effect is about 3 times greater. We ascribe this difference to the features of sample preparing in our case, namely: quenching of samples was conducted in air, i.e. quite slowly. In turn, analyzing in detail the published data [3–8] we found that in all of these studies, quenching of samples was carried out in liquid nitrogen. Below we will argue why we consider this procedure to be very important.

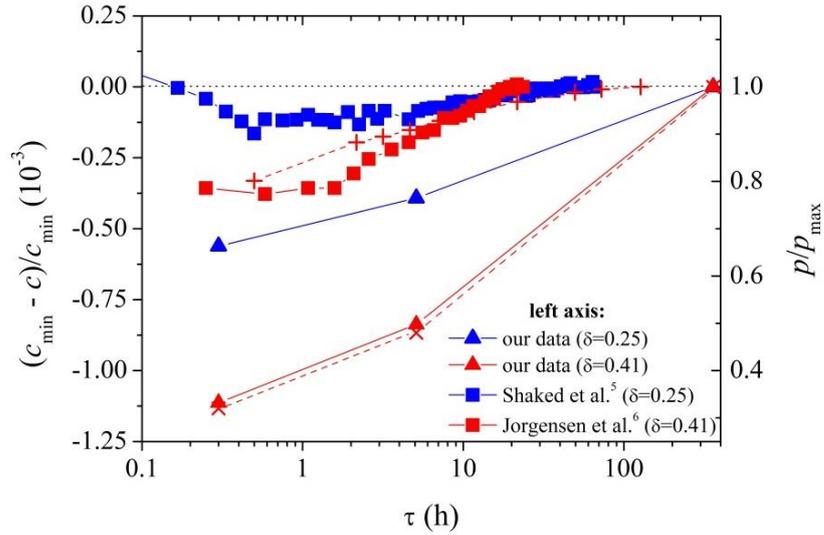

Fig. 4. Comparative analysis of the aging effect found in the present work and in [5, 6]. *The left axis:* relative change in the lattice parameter $c$ (our data are represented by the interpolated values). *The right axis:* relative concentration of holes in the $CuO_2$ plane (see the text in section 4.1); here, data obtained from our XRD measurements (marked with "×" symbols) are compared with the data calculated from the $T_c(τ)$ dependence from [6] using the well-known empirical relationship between $T_c$ and $p$ [37] (marked with "+" symbols).

3.3. *Low-temperature magnetic measurements*

Fig. 5(a) demonstrates the diamagnetic behaviour of the samples under consideration at τ = 15 days; Fig. 5(b) gives values of the superconducting $T_c$ which were defined by the position of the maximum of d$m$/d$T$. As we can see, the superconducting transition widths are relatively narrow for all the $m(T)$ dependences [see Fig. 5(a)] with the exception of *3* and *6*. Note that the transition width in Fig. 5(a) reflects the granular and possibly inhomogeneous structure of the sample where different regions have slightly different response at the same temperature. Meanwhile, especially significant broadening of the transition for the samples $\delta = 0.52$ and 0.76 (see curves *3* and *6*) can be linked to the strong $\delta$-dependence of $T_c$ observed in $YBa_2Cu_3O_{6+\delta}$ in ranges $0.4 < \delta \leq 0.5$ and $0.7 < \delta \leq 0.8$ [see Fig. 5(b)].

It should be noted that $T_c(\delta)$ depicted in Fig. 5(b) is consistent to the data [3, 14, 29, 30] which found $T_c$ to remain practically constant in the range $0.5 < \delta \leq 0.7$.

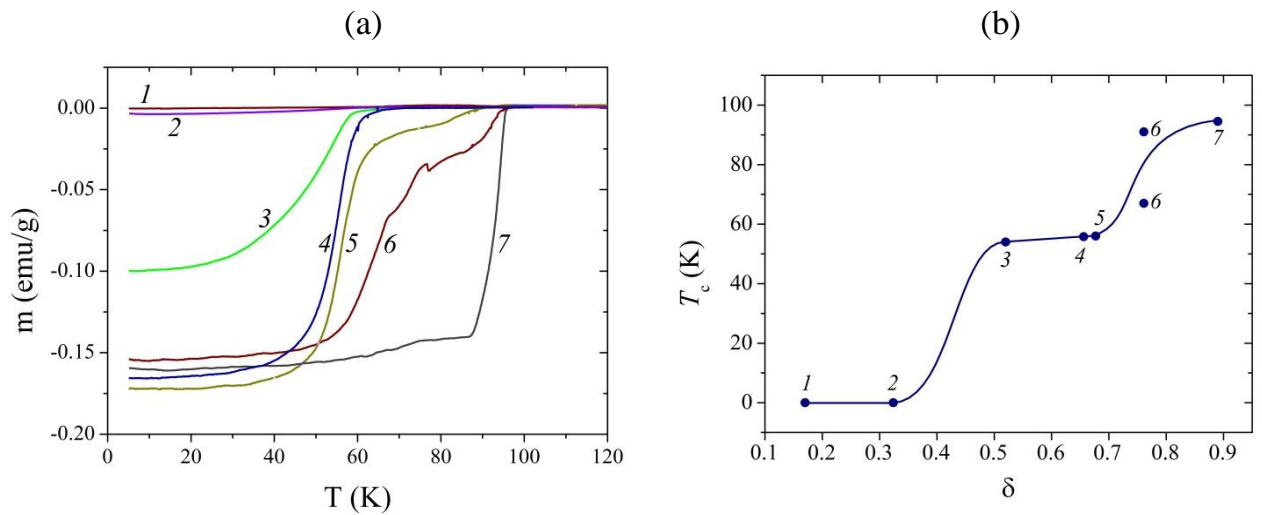

Fig. 5. Low-temperature magnetic studies: (a) The temperature dependence of the magnetic moment of the samples; (b) Superconducting transition temperature as a function of excess oxygen content $\delta$ for $YBa_2Cu_3O_{6+\delta}$. Arabic numerals labeled on the dependence correspond to the curve numbers in fig. (a). The solid curve is a guide to the eye for the data $T_c(\delta)$. In contrast to other samples, the sample with $\delta = 0.76$ shows two very clear superconducting transitions with $T_c = 67$ and 91 K [see curve *6* in Fig. (a)]; in Fig. (b) we have plotted both of these temperatures.

## 4. Discussion

### 4.1. *Modeling of the $c(\delta, \tau)$ dependence. Relationship between the $c(\delta)|_{\tau=const}$ curvature and the hole concentration in $CuO_2$*

Before discussing the $c(\delta, \tau)$ dependence, it is necessary to find out which mechanisms "actuate" compression of the lattice $YBa_2Cu_3O_{6+\delta}$ along the *c*-axis direction when this oxide is saturated with oxygen. It should be noted that a few appropriate mechanisms have already been proposed [29, 31, 32], but they all turn out to be unworkable to describe the $c(\delta, \tau)$ dependence obtained in the present work. So, we offer a new model which allows quantitatively describing the $c(\delta, \tau)$ dependence. We suppose two factors to be responsible for changing the parameter $c$: **(a)** the copper ion radius changing with $\tau$ and with increasing $\delta$ ($r_{Cu^{1+}} \to r_{Cu^{3+}}$), resulting in a decrease in $c$. This factor provides the inversely proportional dependence of $c$ on $\delta$; **(b)** the hole charge transfer from a $CuO_\delta$ plane to two neighboring $CuO_2$ planes giving rise to the nonlinearity of $c(\delta)|_{\tau=const}$. In this case, an increase in $c$ (which becomes more considerable with raising $\delta$, see Fig. 3) results from the electrostatic interaction between those planes.

With regard to factor **(a)**, it should be noted that the dependence of lattice parameters on the ionic radius of cations changing its oxidation state is typical for oxides with various crystal structure (spinels, garnets, perovskites, etc.). For example, heterovalent substitution in $La_{1-x}Ca_xMnO_3$ in the composition range $x \geq 0.15$ results in a decrease in the lattice parameters with increasing $x$ (almost linear relationship); it is entirely the consequence of the decrease in the average radius of manganese cations [33]. Although $r_{Cu^{2+}}$ usually depends on many factors, here we adopt that $r_{Cu^{2+}}$ in $YBa_2Cu_3O_{6+\delta}$ decreases with increasing $z$; it is consistent with the point of view of Shaked *et al.* [5]. In turn, experimental results represented in [34] demonstrate that the transverse load on the $CuO_\delta$ layer "focuses" on the copper ions, "bypassing" the ions of oxygen. This means that strength characteristics as well as the lattice parameter $c$ are entirely determined by the nature of bonds in the structural chains passing through the Cu(1) sites along the $c$ direction [see Fig. 1(a)].

As for factor **(b)**, let us estimate the magnitude of electrostatic interaction between the Cu(1) and Cu(2) planes. According to [30], the distances between different ions in the structural block Cu(2)O(2,3)–Y–Cu(2)O(2,3) are changed slightly with increasing $\delta$. This reflects the fact that yttrium forms robust ionic bonds which are influenced by external electrostatic fields so weakly that such an influence can be neglected. Much more significant variations in the bond lengths are between the $CuO_\delta$ and $CuO_2$ planes and it is the distance that will be the subject of our further consideration. Fig. 6 shows the schematic image of the charge transfer occurring in $YBa_2Cu_3O_{6+\delta}$ where $p$ denotes the hole concentration in $CuO_2$. With increasing $\delta$, the $CuO_\delta$ plane having at $\delta = 0$ a charge per unit cell equal to +1 loses a part of this charge (the loss is equal to $2p$); it appears in the conduction band of the $CuO_2$ planes. If we assume the charge of $CuO_\delta$ initially to be zero (since the potential of the conduction band $\psi \approx \psi_F = 0$) then afterward it increases to $p$. We express the tensile force (per unit area) acting between the two $CuO_2$ planes in Y–$CuO_2$–BaO–$CuO_\delta$–BaO–$CuO_2$–Y structural sandwich as:

$$F = \frac{1}{2\varepsilon_0\sigma^2}\left(\frac{q_1 q_2}{\varepsilon} + q_2 q_2\right) = \frac{3}{2}\left(\varepsilon_{33} \cdot \frac{\Delta l}{l}\right), \qquad (3)$$

where $\varepsilon_0$ is the vacuum permittivity; $\sigma$ is the elementary area on which the force F acts; $q_1$, $q_2$ denote the specific charge of $CuO_\delta$ and $CuO_2$ planes, respectively ($q_i = Q_i \bar{e}$); $\varepsilon$ is the dielectric permittivity which reduces the electric field originated from $q_1$; $\varepsilon_{33}$ is the material elastic constant for the $c$ direction; $\Delta l/l$ denotes the relative tension of

YBa$_2$Cu$_3$O$_{6+\delta}$ along the *c* direction. The constant ε is the polarizability of substance; it, in particular, reflects the ability of metal-oxygen layers contained in layered structures to split under the influence of electric fields. Since fields of different CuO$_2$ layers are mutually nullified at any point of substance, they do not attenuated there. Now, we obtain the final equation for the elastic deformation of YBa$_2$Cu$_3$O$_{6+\delta}$ with account of the constants $\varepsilon_0 = 8.854 \cdot 10^{-12}$ C$^2$/(N·m$^2$), $\sigma^2 = (a \cdot b)^2 \approx 1.5 \cdot 10^{-38}$ m$^2$, $\bar{e} = 1.6 \cdot 10^{-19}$ C, and $\varepsilon_{33} = 186$ GPa [35]:

$$\frac{\Delta l}{l} = 0.345 \frac{Q_1 Q_2}{\varepsilon} + 0.345 \cdot Q_2^2, \qquad Q_2 = p, \qquad Q_1 = 1 - 2p \qquad (4)$$

(It should be noted that the value of the elastic tensor component $\varepsilon_{33}$ for YBa$_2$Cu$_3$O$_{6+\delta}$ is estimated by many researchers to be 61–271 GPa. We have taken the intermediate $\varepsilon_{33}$ value determined by resonant ultrasound spectroscopy). A high polarizability of the BaO layers shown in numerous X-ray studies (e.g., see [30]) allows us to make a rough estimate of ε, which should not be at least less than characteristic of ionic compounds: 6–10. The polarizability data of YBa$_2$Cu$_3$O$_{6+\delta}$ are supported by the presence in the spectrum of characteristic energy losses of the wide region from 2 to 40 eV corresponding to charge density fluctuations [36] (that also implies $\varepsilon \approx 6$–10).

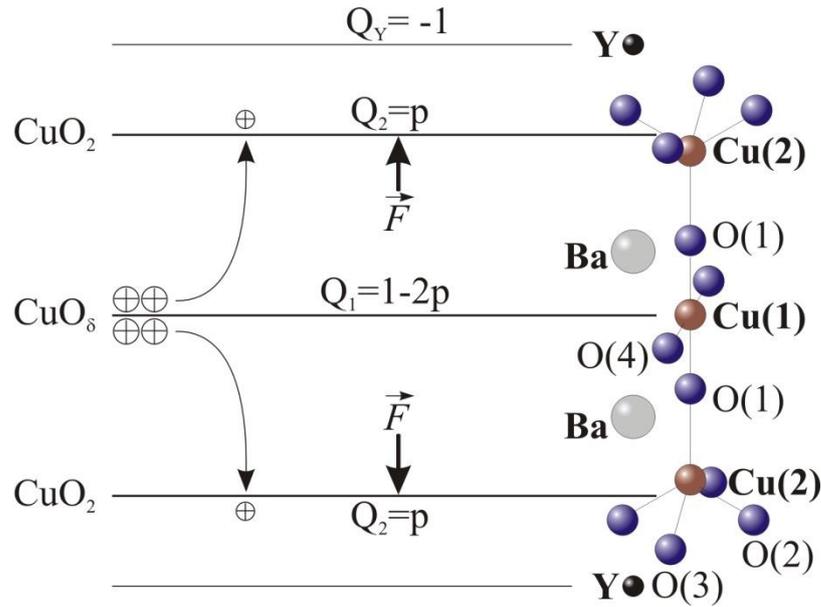

Fig. 6. Electrostatic interaction of structural elements of YBa$_2$Cu$_3$O$_{6+\delta}$ associated with charge transfer. (The charge magnitudes marked on the CuO$_2$ and Y planes do not seem to be relevant to these structural elements. So, herein, these planes should be assumed as CuO and YO$_2$, respectively).

It is easy to notice that terms on the right-hand side of equation (4) are comparable in magnitude, however, the first one contains only a small nonlinear component, so we can consider nonlinear behavior of the $\Delta l/l(p)$ dependence to be entirely determined by its quadratic term. To transform the quadratic term of equation (4) to a dependence

$\Delta c/c(\delta)|_{\tau\to\infty}$ presented as curves *3* in Fig. 7(a), we utilized the data on the relationship between $p$ and $\delta$ from different sources [21, 30, 37]. One should note that in studies [30, 37], where the results of high-precision measurements of the Cu-O bond lengths were obtained, an assessment of the $p$–$\delta$ relationship was made by calculating the bond valence sums for sites of the $CuO_2$ plane. Another assessment [21] was made on the base of an analysis of the $T_c(\delta)$ dependence using empirical equation. Moreover, here we examined the hypothetical case of the relationship between $p$ and $\delta$ represented as $p = k\cdot\delta$, where $k = p_{max} = 0.187$ is the hole concentration at $\delta = 1.0$ reported by Tallon *et al.* [37].

Further, we extract nonlinear part of the $c(\delta)$ experimental dependence "$\tau$ = 15 days" represented in Fig. 3. The method for such extracting is shown in Fig. 7(b), and the resulting data are shown as dataset *3* in Fig. 7(a). As seen, almost all calculated curves *3* describe the experimental data satisfactory. However, the curves which were calculated on the basis of the proportional dependence $p = k\cdot\delta$ and of the dependence $p(\delta)$ obtained by Liang *et al.* [21] (where the $p$–$\delta$ relationship close to proportional is presented) fit very well.

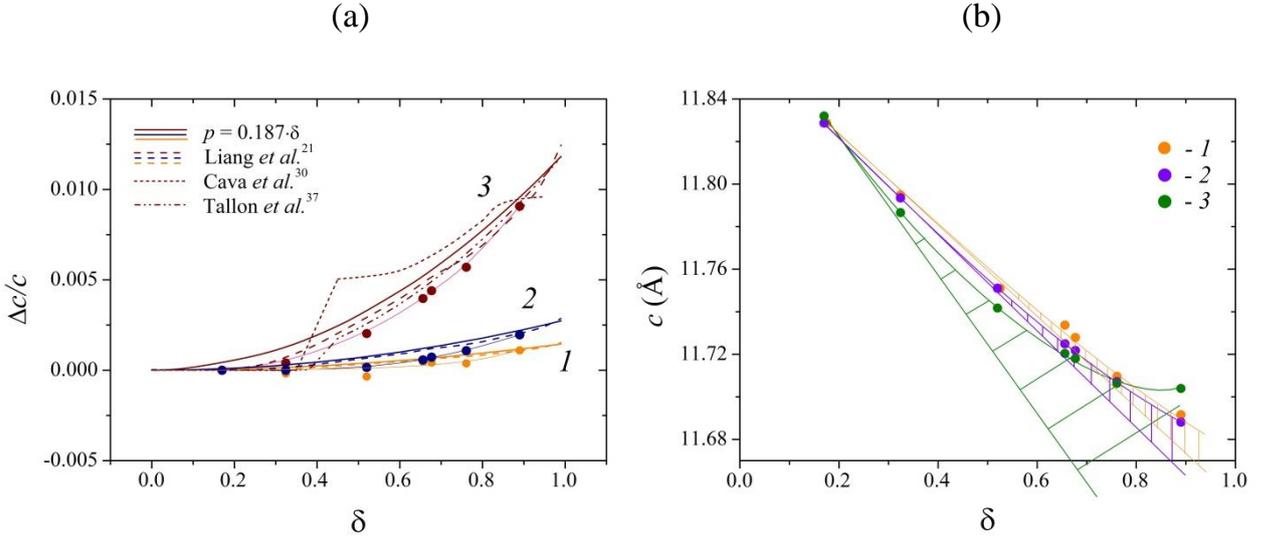

Fig. 7. (a) Description of the nonlinear parts of the $c(\delta, \tau)$ dependence for $\tau$ = 0.3 h (dataset *1*), $\tau$ = 5.1 h (dataset *2*) and $\tau$ = 15 days (dataset *3*) (the last case is assumed to correspond to $\tau\to\infty$) made using equation (4) and data concerning the dependence $p(\delta)$. Experimental curves are marked with dotes and thin curves (guides to the eye); thick lines represent calculations. (b) Extraction of the nonlinear part from experimental data of $c(\delta, \tau)$ for $\tau$ = 0.3 h (dataset *1*), $\tau$ = 5.1 h (dataset *2*) and $\tau$ = 15 days (dataset *3*).

In Fig. 7(a), curves *2* corresponds to the calculations described above, where $Q_2$ is multiplied by a factor of 0.48 and curves *1* – where $Q_2$ is multiplied by a factor of 0.32. These curves are not in bad agreement with experimental datasets *2* and *1*, respectively, and they are given in order to show a possible cause of varying nonlinearity of the $\Delta c/c$ vs $\delta$ dependencies. It is because the charge transfer occurring between the $CuO_\delta$ and $CuO_2$ planes does not have time to come to equilibrium during 0.3 and 5 h but the equi-librium turns out to be achieved after storing the samples for 15 days at RT. Fig. 4 shows, in addition to the data concerning the aging effect in terms of $c(\tau)|_{\delta=const}$, the data on the hole concentration in $CuO_2$ [$p(\tau)$] found above as well as calculated on the basis of a $T_c(\tau)$ dependence from [6]. The comparison of different types of dependencies [$c(\tau)$ and $p(\tau)$] in Fig. 4 clearly shows a good correlation between them. This indicates that parameter $c$ adequately reflects the hole concentration in the $CuO_2$ plane and can serve as a tool to evaluate $p$.

Finally, it should be noted that the course of the $a(\delta, \tau)$ dependence for $\tau = 15$ days in the ortho-phase area (see Fig. 3), which is very similar to the behavior of $c(\delta, \tau)$, can be explained similarly: in terms of the copper ionic radius and the electrostatic interaction between these ions within the $CuO_\delta$ plane (which is not screened by oxygen in the direction $a$ only).

4.2.  *New view on the nature of aging effects in $YBa_2Cu_3O_{6+\delta}$*

As already mentioned in the Introduction, the aging effects observed on quenched $YBa_2Cu_3O_{6+\delta}$ samples usually is assigned to the processes of oxygen ordering occurring at RT for a long time [4–8]. At the same time, the existence of this relationship can be supported or refuted by using the results of the previous section. Indeed, if the "ordering factor" determines the hole concentration in the system, this should be reflected in the $c(\delta, \tau)$ dependence. Its particular cases corresponding to large $\tau$ (when the formation of low-temperature phases is finished in the system) should have singularities at the structural-phase transition points: $\delta = 0.32, 0.63, 0.77$, etc. [13, 16, 18] (for tetra→орто-II, орто-II→орто-III, орто-III→орто-I, etc., respectively). With increasing $\tau$, these singularities should become more noticeable. However, in fact, we see a completely opposite picture: the $c(\delta, \tau)$ dependence shows correlations with the low-temperature structural phase diagram of $YBa_2Cu_3O_{6+\delta}$ only in a certain neighbourhood of $\tau = 0$ (see Fig. 3). Out of this $\tau$, all correlations completely disappear. It is also interesting that there are no breaks in $c(\delta)$ curves at $\delta = 0.5$, suggesting sequential oxidation of the copper ions from $Cu^+$ to $Cu^{2+}$ at the beginning and only then to $Cu^{3+}$. This behaviour implies that oxygen structure is not involved in the process of hole formation. When $\tau$ is large ($\tau\to\infty$), all neighboring O(4) and Cu(1) ions are involved in the formation of holes owing to reaction (1)[*]. Completed reaction (1) gives rise to the proportional dependence of $p(\delta)$ (assuming an equal contribution of holes in the $CuO_\delta$ plane to the charge transfer between $CuO_\delta$ and $CuO_2$ planes). It is the dependence that allows describing our experimental data $\Delta c/c(\delta)|_{\tau\to\infty}$ better than in the case of other available $p(\delta)$ dependences [see Fig. 7(a)].

---

[*] However, that is possible only if to assume that the birth of holes can occur not only on the square-planar-coordinated copper as it is believed, but also in some other cases, for example, on a three-coordinated copper [see chain fragments of type II in Fig. 1(b)] with forming the double bond Cu=O.

However, the universality of reaction (1) revives the old question, why the hole concentration in $YBa_2Cu_3O_{6+\delta}$ is significantly reduced just after quenching and grows for a long time later on.

XPS data represented in Fig. 2 allow relating the changes in the hole concentration $p$ to the changes observed in the O $1s$ spectrum. The latter point out that for a while after quenching, holes in the $CuO_\delta$ plane stay on oxygen (this electronic configuration is usually denoted as Cu $3d^9L^{-1}$) but in a day they turn out on copper (Cu $3d^8$ configuration). The long-term charge transfer occurring between adjacent O(4) and Cu(1) sites indicates that the holes placed on oxygen are in a strongly localized state. From the standpoint of pd-model, that state could point to a small magnitude of the charge transfer matrix element for the Cu(1)–O(4) pairs. Or, it could mean, in fact, the absence of holes on the O $2p$ orbitals (this could be in the case of formation of the peroxide oxygen $(O_2)^{-2}$ from adjacent $O^-$ ions). Anyway when the $3d^9L^{-1}$ configuration takes place, the charge transfer between $CuO_\delta$ and $CuO_2$ planes, implemented over the chain O(4)–Cu(1)–O(1)–Cu(2), cannot occur more quickly than it occurs between elements of the first link. Hence, its minimum duration should be about a day. It is the duration that is observed for the aging effects we are known from most studies [4–8].

The possibility for existence of holes on oxygen sites in $YBa_2Cu_3O_{6+\delta}$ was justified in a number of works. Thus, for example, studying the equilibrium between $YBa_2Cu_3O_{6+\delta}$ and gas phase, Verwey [38] came to the necessity to consider the high-temperature redox process of this oxide as follows:

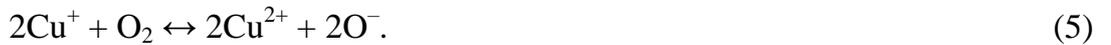

$$2Cu^+ + O_2 \leftrightarrow 2Cu^{2+} + 2O^-. \tag{5}$$

At the same time, numerous experimental evidences for copper ions existing in $YBa_2Cu_3O_{6+\delta}$ as $Cu^{3+}$ at RT (for example, [34, 39, 40]) lead us to the assumption that oxygen hole-localization in the pair Cu–O exists at high temperature, but copper hole-localization – at RT (by analogy with [15]). The transition between these two states of the system can be represented by the electronic reaction:

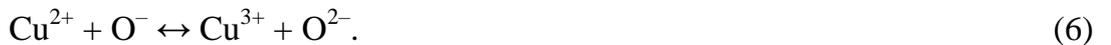

$$Cu^{2+} + O^- \leftrightarrow Cu^{3+} + O^{2-}. \tag{6}$$

A temperature $T_{tr}$ at which reaction (6) is activated is proposed to be about 600 K (as in [15]).

We believe it is reaction (6) that is the process observed for quenched samples at RT. Its kinetic features can be explained by assuming that while $YBa_2Cu_3O_{6+\delta}$ passing the temperature range near $T_{tr}$ slowly, the oxygen $O^-$ has enough time to form the complexes $(O_2)^{-2}$ being metastable at RT.

## 5. Conclusions

The aging effects occurring in $YBa_2Cu_3O_{6+\delta}$ have been experimentally studied in a large number of works. Typically, the changes in $T_c$ and $c$ observed after quenching

$YBa_2Cu_3O_{6+\delta}$ are explained by the ordering processes proceeding in the incomplete oxygen sublattice of the $CuO_\delta$ plane at RT. It is generally accepted that $T_c$ is greater, when the part of the square-planar coordinated copper, as a source of the superconducting holes, is higher.

In the present investigation, however, a new way to quantify the concentration of holes $p$ in $YBa_2Cu_3O_{6+\delta}$ has been developed. It has been found a link between $p$ and the $c(\delta)$ dependence. Detailed analysis of the experimental dependence $c(\delta, \tau)$, where $\tau$ is the time elapsed after quenching the sample, allows deducing that oxygen ordering of the $CuO_\delta$ plane don't influence on $p$ opposite to general opinion. In turn, XPS study provides experimental evidence that for a while after quenching $YBa_2Cu_3O_{6+\delta}$, its holes are located on the oxygen sites O(4), that corresponds to a weak charge transfer from $CuO_\delta$ to $CuO_2$. Over time, however, the hole charge shifts to the copper sites Cu(1), that leads to enhancement of the charge transfer between the planes. Thus, XPS study explains the nature of aging effects regardless of oxygen ordering.

**Acknowledgements**